# Study of soft/hard bimagnetic CoFe$_2$/CoFe$_2$O$_4$ nanocomposite

E. H. Coldebella[1], E. F. Chagas[1], A. P. Albuquerque[1], R. J. Prado[1], M. Alzamora[2], and E. Baggio-Saitovitch[3]

[1]Instituto de Física, Universidade Federal de Mato Grosso, 78060-900 Cuiabá-MT, Brazil.
[2]Campus Duque de Caxias, Universidade Federal do Rio de Janeiro, 25265-008 RJ, Brazil
[3]Centro Brasileiro de Pesquisas Físicas, Rua Xavier Sigaud, 150 Urca Rio de Janeiro, Brazil

## Abstract

We report an experimental study of the bimagnetic nanocomposites CoFe$_2$/CoFe$_2$O$_4$. The precursor material, CoFe$_2$O$_4$ was prepared using the conventional stoichiometric combustion method. The nanocomposite CoFe$_2$/CoFe$_2$O$_4$ was obtained by total reduction of CoFe$_2$O$_4$ using a thermal treatment at 350°C in H$_2$ atmospheres following a partial oxidation in O$_2$ atmospheres at 380°C during 120; 30; 15, 10, and 5 min. The X-ray diffraction and Mossbauer spectroscopy confirmed the formation the material CoFe$_2$/CoFe$_2$O$_4$

The magnetic hysteresis with different saturation magnetization confirms the formation of the CoFe$_2$/CoFe$_2$O$_4$ with different content of CoFe$_2$O$_4$. The high energy milling to the precursor material increase the coercivity from 1.0 to 3.3 kOe, however the same effect was not observed to the CoFe$_2$/CoFe$_2$O$_4$ material.

Cobalt ferrite; nanocomposite; nanomagnetism

## INTRODUCTION

The cobalt ferrite (CoFe$_2$O$_4$) belong to a group of metal ferrite like $M^{2+}Fe_2^{3+}O_4^{2-}$ that had attracted considerable interest due potentials to several applications [1-6]. Among the hard ferrites the CoFe$_2$O$_4$ presents interesting characteristics, such as electrical insulation, chemical stability, high magnetic-elastic effect [7], moderate saturation magnetization, thermal chemical reduction, and high coercivity. Due to these characteristics, the cobalt ferrite is a promising material for several technological applications.

Particularly for permanent magnets applications coercivity ($H_C$) and saturation magnetization ($M_S$) are fundamentals. Both quantities composes the figure of merit in a hard magnetic material, the energy product (BH)$_{max}$, this quantity, in a simplified way, gives an idea of the amount of energy magnetic that can be stored in the material. Consequently, to optimize the use CoFe$_2$O$_4$ for this type of application, it is convenient to increase their $M_S$ and $H_C$.

Several works reports successfully methods used to increase the $H_C$ of cobalt ferrite, such as thermal annealing [8], capping [9] and mechanical milling treatment [10, 11, 12].

The exchange spring effect has been used to increase the $M_S$ of some nanomaterials [13-17]. This effect was observed for the first time in 1989 by Coehoorn et al. [13] and explained in 1991 by Kneller and Hawig [17], who argue that, under certain conditions, hard and soft magnetic materials may present exchange coupling. The exchange spring magnets combine in a single material (nanocomposite), the high $H_C$ from the hard material with the high $M_S$ from the soft material, increasing substantially the product $(BH)_{max}$ when compared with any one of the individual phases of the nanocomposite.

Fortunately, $CoFe_2O_4$ is also a promising material for obtaining optimized exchange-spring magnets [14-16] due the easy process for obtention of $CoFe_2$ by reduction of $CoFe_2O_4$. The $CoFe_2$ is a soft ferromagnetic material with high $M_S$ (bigger than 200 emu/g). Two methods have been used to this purpose: reduction with hydrogen atmosphere at 300ºC ($CoFe_2O_4 + 4H_2 \rightarrow CoFe_2 + 4H_2O$) [14] and reduction with carbon at 900ºC ($CoFe_2O_4 + 2C \rightarrow CoFe_2 + 2CO_2$) [16]. However, the first method presents important advantages, due the relatively low process temperature, that prevent coalescence between the nanoparticles, and provide better control of the thickness of the reduced material. By the other way, the $CoFe_2$ also presents an interesting and useful characteristic, the oxidation transforming into $CoFe_2O_4$ ($CoFe_2 + 2O_2 \rightarrow CoFe_2O_4$). This property was used by Scheffe et al. to produce hydrogen [6].

In 2008, Cabral et al. successfully fabricated $CoFe_2O_4$(core)-$CoFe_2$(shell) nanocomposite by reducing $CoFe_2O_4$ nanoparticles in $H_2$ [14]. Leite et al. obtained the same nanocomposite reducing $CoFe_2O_4$ with carbon [16], however in the both cases with small values of $(BH)_{max}$, as compared with high-Hc $CoFe_2O_4$ [10-12]. Thus, more studies trying to improve the energy product $(BH)_{max}$ are necessary. Moreover, only few works investigated the system $CoFe_2$(core)-$CoFe_2O_4$(shell) [18], reasons why we consider that this kind of studies deserves attention. Due to these reasons, in this work we investigated the nanocomposites $CoFe_2$(core)-$CoFe_2O_4$(shell) with different content of $CoFe_2O_4$ and tried to increase the coercivity of the nanocomposite by milling, how occurs to pure $CoFe_2O_4$ [10].

## EXPERIMENTAL

The precursor material, $CoFe_2O_4$, was prepared using a conventional combustion method [19]. High-purity (99.9%) raw compounds were used. Cobalt nitrate ($Co(No_3)_2 6H_2O$) and iron nitrate ($Fe(NO_3)_3 9H_2O$) were dissolved in 450 ml of distilled water in a ratio corresponding to the selected final composition. Glycine ($C_2H_2NO_2$) was added in a proportion of one and half moles per mole of metal atoms, and the pH of the solution was adjusted with ammonium hydroxide in the range of 3 to 7. The pH was tuned as closely as possible to 7 avoiding precipitation. The resulting solution was concentrated by evaporation using a hot plate at 300ºC until a viscous gel was obtained. This hot gel finally burnt out as a result of a vigorous exothermic reaction. The system remained homogeneous throughout the entire process and no precipitation was observed. Finally, the as-reacted material was calcined in air at 700ºC for 2 h in order to remove the organic residues.

The process to obtain the nanocomposite $CoFe_2$-$CoFe_2O_4$ can be described in two distinct parts: firstly, the $CoFe_2O_4$ the precursor material was submitted to a thermal treatment at 350ºC in a tubular furnace at hydrogen atmosphere for 5 h. The process transforms all cobalt ferrite in $CoFe_2$ according with the chemical equation:

$$CoFe_2O_4 + 4H_2 \underset{\Delta}{\rightarrow} CoFe_2 + 4H_2O \quad (1)$$

The symbol $\Delta$ indicates that thermal energy is necessary in the process.

The second step is to transform part of the $CoFe_2$ in a $CoFe_2O_4$ by oxidation, using the same tubular furnace working at 380 ºC with pure oxygen atmosphere, but different processing times to tune the thickness/content of the cobalt ferrite shell. The oxidation process can be described by:

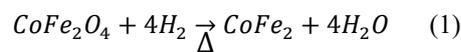

$$CoFe_2 + 2O_2 \underset{\Delta}{\rightarrow} CoFe_2O_4 \quad (2)$$

The all process of $CoFe_2/CoFe_2O_4$ preparation is illustrated by figure 1.

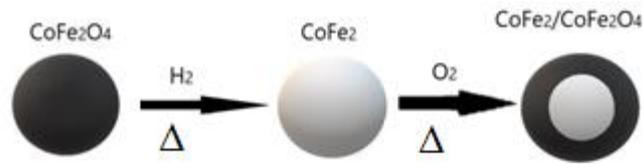

Figure 1 – schematic picture of the reduction and oxidation processes.

We prepared five samples of the nanocomposite $CoFe_2/CoFe_2O_4$ with different content of $CoFe_2O_4$ obtained by 120, 30, 15, 10, and 5 minutes of thermal treatment described by the chemical reaction (2). These samples were labeling respectively as: CF-CFO_120; CF-CFO_30; CF-CFO_15; CF-CFO_10, and CF-CFO_5.

Finally, a Spex 8000 high-energy mechanical ball miller with 6 mm diameter zirconia balls, was employed for milling processing for four samples (pure $CoFe_2O_4$; CF-CFO_120; CF-CFO_30; CF-CFO_15), aiming exclusively to increase their coercivity. The processing time was 1.5 h for all samples with ball/sample mass ratio of about 1/9. Detailed milling conditions are described in Ponce et al. [10].

The crystalline phases of the nanocomposite were identified by X-ray diffraction (XRD), using a Shimadzu XRD-6000 diffractometer installed at the *Laboratório Multiusuário de Técnicas Analíticas* (LAMUTA/ UFMT–Cuiabá-MT– Brazil). It is equipped with graphite monochromator and conventional Cu tube (0.154178 nm), and works at 1.2 kW (40 kV, 30 mA), using the Bragg-Brentano geometry. Magnetic measurements (hysteresis loops at 300 and 50 K) were carried out using a vibrating sample magnetometer (VSM) model VersaLab Quantum Design, and the $^{57}Fe$ Mössbauer spectroscopy experiments were performed at room temperature in transmission geometry with the Co-57 in Rh-matrix source moving in a sinusoidal mode. Isomer shifts (δ) are reported relative to α-Fe at room temperature, both installed at CBPF, Rio de Janeiro RJ–Brazil.

## RESULTS AND DISCUSSION

The figure 2 presents the XRD patterns of the precursor material ($CoFe_2O_4$) before (CFO sample) and after (CF sample) the 5 h thermal treatment in a reductive $H_2$ atmosphere, as indicated by chemical reaction (1). These patterns suggest the total transformation of $CoFe_2O_4$ (fig. 2*a*) into $CoFe_2$ (fig. 2*b*).

Using the Scherrer equation to the more intense XRD peak, we obtained 52 nm to the mean size of the crystallite for the sample CFO.

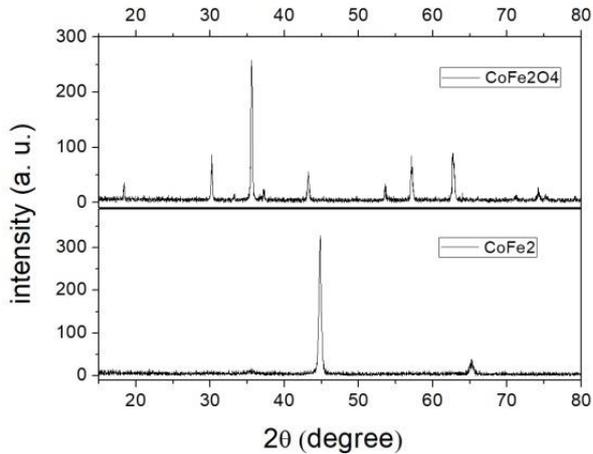

Figure 2 - XRD patterns of the precursor material CoFe$_2$O$_4$ (CFO sample) and CoFe$_2$ obtained after its complete reduction in H$_2$ atmosphere (CF sample).

We also performed the Mossbauer spectroscopy to evaluate the chemical reduction process, however, to understand the results, it is convenient to know the crystalline structure of CoFe$_2$O$_4$. The cobalt ferrite has an inverse spinel structure, which can be described by [20]: $(Co_{1-i}^{2+}Fe_i^{3+})_A[Co_i^{2+}Fe_{2-i}^{3+}]_B O_4^{2-}$, in this representation, the round and the square brackets indicate $A$ (tetrahedral) and $B$ (octahedral) sites, respectively, and $i$ (the degree of inversion) describes the fraction of the tetrahedral sites occupied by Fe$^{3+}$ cations. The ideal inverse spinel structure has $i = 1$ and a mixed spinel structure present $i$ values between 0 and 1. Generally, the cobalt ferrite is in the second case, and, consequently, Mössbauer spectra can be typically analyzed on the basis of one subspectrum associated to Fe$^{3+}$ in tetrahedral sites and one (or more) subspectra arising from Fe$^{3+}$ in octahedral sites [21]. In the simplest analysis two sextets are expected. In the other hand, a complete reduction of the CoFe$_2$O$_4$ (CFO sample) to CoFe$_2$ (CF sample) transform the spinel to a body center structure, which give an only one position to Fe$^{+3}$. The Mössbauer spectroscopy is powerful tools to study local electronic structure and can distinguish these different scenarios.

In figure 3 $^{57}$Fe Mössbauer spectra of CoFe$_2$ after first reaction (described by chemical equation 1) are shown. The spectrum exhibited a magnetic hyperfine structure consisting mainly of a six-line (S) pattern with an isomer shift (δ) of 0.032 ± 0.006 mm/s and an observed hyperfine magnetic field (B$_{hf}$) of 36.1 ± 0.5 T, corresponding to the 96 % of the total area, could be associated to the CoFe$_2$ phase. A broad six-line pattern, corresponding to 4% of the area was fit assuming a linear correlation of hyperfine field with isomeric shift. The medium value of the B$_{hf}$ obtained was 48.1 ± 1 T and δ was 0.310 ± 0.008 mm/s. Although it is difficult to resolve the individual contributions of the A and B sites, we associated the distribution to CoFe$_2$O$_4$ phase. The result revels that chemical reaction described by eq. (1) occurs, but the reduction of the material was not totally achieved, suggesting the presence of about 4% cobalt ferrite after reduction.

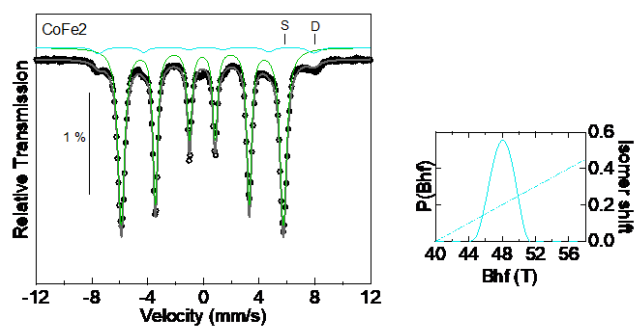

Figure 3- Mössbauer spectrum of CF sample and theoretical fit of the data.

With respect to the transformation of $CoFe_2O_4$ into $CoFe_2$, the Mossbauer results are in agreement with XRD measurements, however the result obtained in this work indicate that Mossbauer spectroscopy is much more sensitive than XRD technique to observe small amounts of cobalt ferrite, which is understandable since Mossbauer spectroscopy is a local atomic technique (sensitive to the oxidation number of the Fe atoms, a property defined basically by their first neighbors) while XRD is sensitive of crystalline atomic structures.

It is well known that the magnetic behavior of $CoFe_2O_4$ and $CoFe_2$ are quite different, the first one is a hard ferrimagnetic material with $H_C$ between 0.70 up to 9.5 kOe and moderate $M_S$ (about 70 emu/g) while the second is a soft ferromagnetic material with small $H_C$ and very high $M_S$ (higher than 200 emu/g). The figure 4 presents the room temperature hysteresis to the samples CFO and CF, there is a clear difference of magnetic behavior. The $M_S$ value obtained for the CF sample (208 emu/g) is about three times bigger than the value obtained for the samples CFO (77 emu/g), and the coercivity of CF ($H_C$ = 0.3 kOe) is three times smaller than the value obtained for the CFO sample (0.9 kOe). These results are consistent with the expected magnetic behavior of $CoFe_2O_4$ and $CoFe_2$, furthermore, the magnetic behaviors are in complete agreement with the XRD and Mossbauer spectroscopy data, which indicate an almost complete conversion of $CoFe_2O_4$ in $CoFe_2$ after reduction in $H_2$ atmosphere.

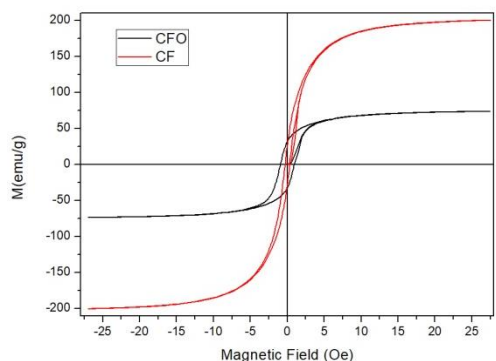

Figure 4 – Magnetic hysteresis of CFO and CF samples at room temperature.

The bimagnetic $CoFe_2$-$CoFe_2O_4$ nanocomposites were prepared using CF sample and the oxidation process described by the chemical reaction (2), being that the thickness of the shell depends on the duration of thermal treatment in $O_2$ atmosphere. The XRD patterns of samples CF-CFO_120; CF-

CFO_30; CF-CFO_15; CF-CFO_5 presents the characteristic peak of $CoFe_2$ at approximately 45º in 2θ (* in fig. 5), however the peak at 65º in 2θ (see figure 2b) is difficult to identify due its small intensity. The characteristic peaks of the $CoFe_2O_4$ shown in figure 2a are also visible, confirming the preparation of the bimagnetic material $CoFe_2$-$CoFe_2O_4$. The large background might be caused by fluorescence of Fe atoms. This effect difficult to obtain the mean size of crystallite from Scherrer equation.

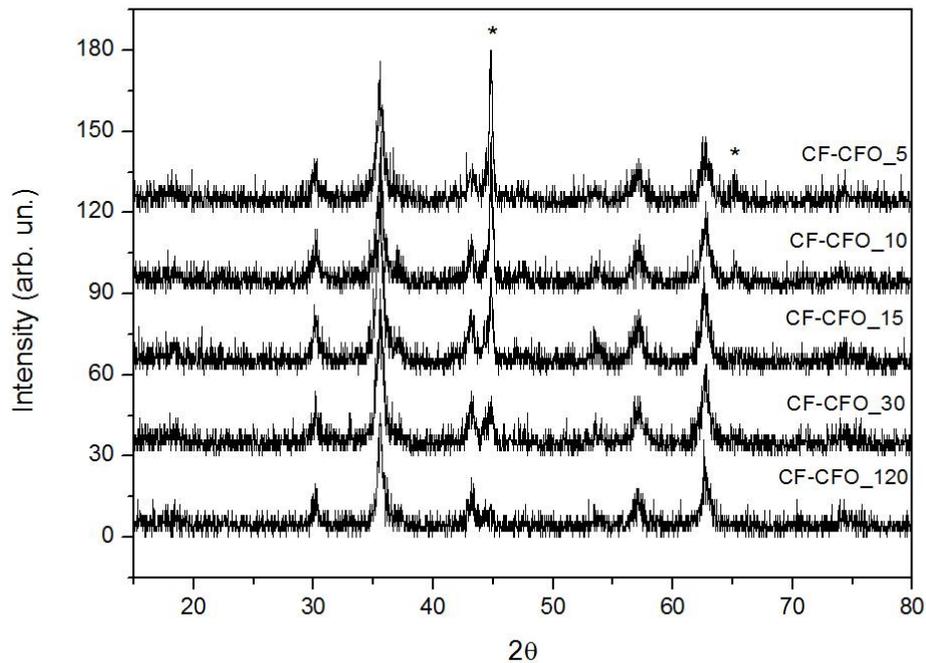

Figure 5 - The XRD patterns to the samples CF-CFO_120; CF-CFO_30; CF-CFO_15; CF-CFO_10; CF-CFO_5.

Mössbauer measurements were also performed in samples CF-CFO_5 and CF-CFO_10. The Mössbauer spectra of both samples show a clear magnetic pattern and were fitted with three magnetic sextets (Fig. 6), the hyperfine parameters for the three sextets are shown in the table 1. The hyperfine parameters of the site S are typical of the $CoFe_2$ phase and the sites SA and SB are attributed to the tetrahedral and octahedral sites of the spinel structure cobalt ferrite, respectively. As we can see from table 1, the absorption area of the site SA and SB increases significatively when compared with that corresponding to sample without heat treatment showing that the oxidation process was successfully achieved with only 5 minutes of heat treatment. The amount of $CoFe_2O_4$ drastically increased from 4 to 72% in 5 minutes.

Table 1 - Hyperfine parameters obtained from the fit of Mossbauer spectra at room temperature to the samples CF-CFO_5 and CF-CFO_10.

| Sample | S | | | SA | | | SB | | |
|---|---|---|---|---|---|---|---|---|---|
| | δ ± 0.003 mm/s | Bhf ± 0.5 T | Are ± 1 % | δ ± 0.03 mm/s | Bhf ± 0.5 T | Are ± 1 % | δ ± 0.03 mm/s | Bhf± 0.5 T | Are ± 1 % |
| CF-CFO_5 | 0.028 | 36.0 | 26 | 0.532 | 45.7 | 23 | 0.334 | 49.3 | 50 |
| CF-CFO_10 | 0.028 | 36.2 | 28 | 0.536 | 45.8 | 25 | 0.361 | 49.2 | 47 |

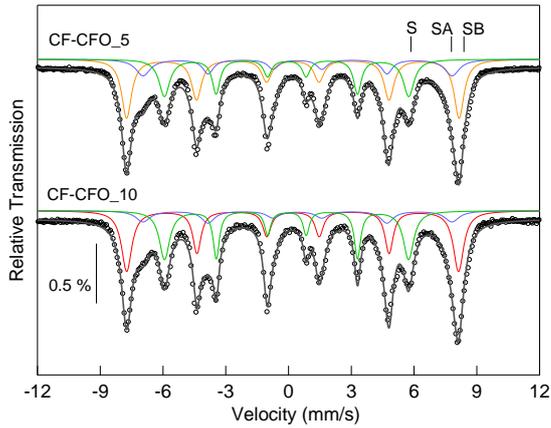

Figure 6 – Mossbauer spectra and fittings of samples CF-CFO_5 and CF-CFO_10.

The magnetic hysteresis curves (fig. 7) of the samples CF-CFO_120; CF-CFO_30; CF-CFO_15; CF-CFO_5 are in agreement with the XRD and Mossbauer spectroscopy results. The differences in $M_S$ between the samples observed in the hysteresis curves confirm the formation of the $CoFe_2/CoFe_2O_4$ composite with different contents of $CoFe_2O_4$. As expected, the sample treated during more time presented a smaller saturation magnetization, this result is consistent with more content of cobalt ferrite in the nanocomposite (sample CF-CFO_120). All samples follow this tendency, thus, the sample with biggest $M_S$ was the CF-CFO_5. On the other hand, the coercivity increases with the content of $CoFe_2O_4$, that also is an expected effect due the magnetic hardness of the cobalt ferrite.

The single magnetic behavior in the hysteresis (no kinks in the curve) indicates the magnetic coupling between $CoFe_2O_4$ and $CoFe_2$, furthermore this behavior is expected only to materials with nanometric scale [22].

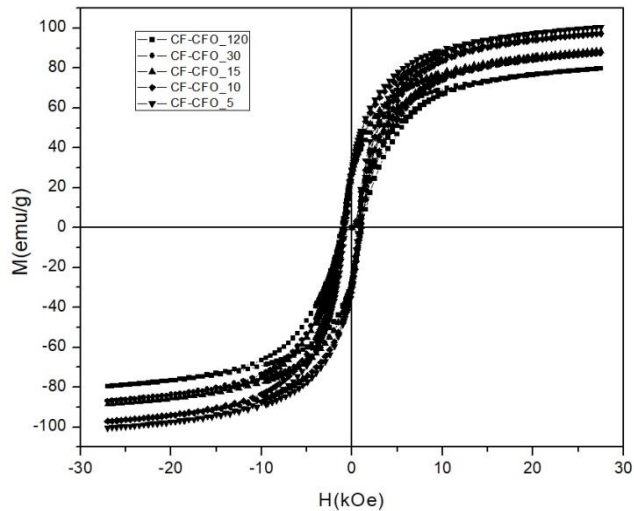

Figure 7- Magnetic hysteresis of the nanocomposites: CF-CFO_120; CF-CFO_30; CF-CFO_15; CF-CFO_5.

Using the saturation magnetization of the nanocomposite sample and considering the $M_S$ expected to pure inverse spinel cobalt ferrite (71.4 emu/g) and pure $CoFe_2$ (230 emu/g) [23] was possible to estimate the content of $CoFe_2O_4$ in each sample employing a simple calculation. In the table 2 the content of $CoFe_2O_4$ obtained from $M_S$ and Mossbauer spectroscopy data are compared. It is possible to observe a reasonable concordance between the content of $CoFe_2O_4$ obtained from Mossbauer spectroscopy and from the magnetic data to the samples CF-CFO_5 and CF-CFO_10.

Table 2 - $CoFe_2O_4$ content determined by Mossbauer spectroscopy and magnetic measurements to the different nanocomposites.

| Sample | $CoFe_2O_4$ content from | |
|---|---|---|
| | Mossbauer (%) | $M_S$ (%) |
| CF | 4 | 14 |
| CF-CFO_5 | 73 | 77 |
| CF-CFO_10 | 72 | 79 |
| CF-CFO_15 | - | 84 |
| CF-CFO_30 | - | 86 |
| CF-CFO_120 | - | 90 |

The magnetic parameters from the magnetic hysteresis are summarized in the table 3. The $M_S$ values were obtained extrapolating to zero the M versus 1/H plot. One can see that there is a significant increase of $M_R/M_S$ ratio with the increase of the $CoFe_2O_4$ content to all samples CF-CFO, as compared with the CF sample, however, the value of $M_R/M_S$ remained smaller than obtained to CFO sample. The values of product $(BH)_{max}$ for all nanocomposites were smaller than obtained for the CFO sample, despite bigger than obtained for CF sample ($CoFe_2$).

Table 3 – Magnetic parameters:

| Sample | Ms (emu/g) | $H_C$ (kOe) | $M_R$(emu/g) | $M_R/M_S$ | $(BH)_{max}$ (MGOe) |
|---|---|---|---|---|---|
| CFO | 77 | 1.0 | 32 | 0.42 | 0,41 |
| CF-CFO_120 | 87 | 1.0 | 27 | 0.3 | 0,36 |
| CF-CFO_30 | 94 | 0.97 | 28 | 0.3 | 0,34 |
| CF-CFO_15 | 96 | 0.97 | 28 | 0.29 | 0,34 |
| CF-CFO_10 | 105 | 0.87 | 29 | 0.28 | 0,38 |
| CF-CFO_5 | 108 | 0.78 | 29 | 0.27 | 0,34 |
| CF | 209 | 0.30 | 28 | 0.13 | 0,18 |

Our results indicate that the nanocomposites present significant increase of $M_S$ as compared with pure cobalt ferrite, however, they present small values of $H_C$ as compared with values obtained, for example, by Leite et al. [16] (1,9 kOe) and by B. H. Liu and J. Ding [11] (5.1 kOe). The small values of coercivity resulted in an undesirable behavior to hard magnetic application, small values of $(BH)_{max}$ (see table 2), for comparison Galizia et al. [12] and Kumar et al. [24] obtained respectively 2.16 and 2.41 MGOe for very hard cobalt ferrite.

In order to increase the coercivity we performed a high energy milling process in the precursor material (CFO sample) and in the samples CF-CFO_120, CF-CFO_30 and CF-CFO_15. The figure 8$a$ compares the hysteresis of $CoFe_2O_4$ (CFO sample) before and after the milling process. The figure reveals an enormous increase in $H_C$. This effect was explained by the references [10, 11, 19], it is associated with increase of strain and density of structural defects. Moreover, was observed a decrease in $M_S$, this effect is related to the decrease of the average crystallite size and consequent increase of contribution of canted magnetic moment of surface magnetic cations. The $(BH)_{max}$ also increase from 0.41 to 0.60 MGOe, even with the decrease of the saturation magnetization.

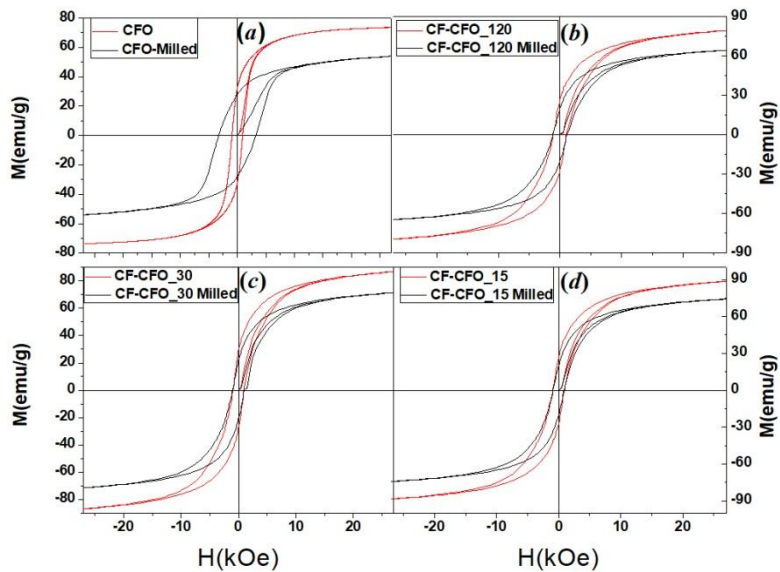

Figure 8 – Magnetic hysteresis before and after the milling process to sample ($a$) CFO; (b) CF-CFO_120, ($c$) CF-CFO_30 and ($d$) CF-CFO_15.

Nevertheless, to samples CF-CFO_120, CF-CFO_30, and CF-CFO_15 the milling process did not result the same effect to the coercivity. To the sample CF_CFO_120 the $H_C$ increased from 1.0 to 1.1 kOe; to the sample CF-CFO_30 the $H_C$ remained 1.0 kOe after the milling process and to the sample CF-CFO_5 was observed a slightly decrease from 0.97 to 0.94kOe. These effects for the coercivity associated with the decrease of $M_S$ for all milled samples caused a significant decrease of $(BH)_{max}$ for these samples (see table 4).

Table 4 – Magnetic parameters to the milled samples.

|  | Ms (emu/g) | $H_C$ (kOe) | $M_R$(emu/g) | $M_R/M_S$ | $(BH)_{max}$ (MGOe) |
|---|---|---|---|---|---|
| CFO_Milled | 59 | 3.3 | 28 | 0.47 | 0.60 |
| CF-CFO_120_Milled | 70 | 1.1 | 20 | 0.29 | 0.21 |
| CF-CFO_30_Milled | 78 | 1.0 | 21 | 0.27 | 0.24 |
| CF-CFO_15_Milled | 81 | 0.94 | 20 | 0.25 | 0.20 |

There is an important difference from the $CoFe_2O_4$ of CFO sample and $CoFe_2O_4$ of the samples CF-CFO_120, CF-CFO_30, and CF-CFO_15, the first was obtained from the conventional combustion method and second obtained from the oxidation of $CoFe_2$. For this reason is expected structural differences between them, indeed, more investigations are needed to understand this unexpected behavior to the nanocomposite.

## CONCLUSION

We prepared the $CoFe_2/CoFe_2O_4$ nanocomposite successfully, confirmed by XRD, Mossbauer spectroscopy and magnetic hysteresis. We obtained this material using the chemical reduction property of cobalt ferrite associate with oxidation property of iron cobalt.

The one phase behavior of magnetic hysteresis obtained to the $CoFe_2/CoFe_2O_4$ suggests the magnetic coupling between $CoFe_2$ and $CoFe_2O_4$ phases. However, the small values of $M_R/M_S$ ratio limit the hard magnetic applications.

The precursor material $CoFe_2O_4$ responded to the high energy milling, as expected, increasing substantially the coercivity and decreasing the $M_S$, however to the samples $CoFe_2/CoFe_2O_4$ (including sample with high concentration of $CoFe_2O_4$) the results were quite different, no significant changes was observed to the coercivity, suggesting that the structural differences are fundamental to the effect of change in $H_C$. However, this effect deserves more investigation.